\documentclass[twocolumn,prl,showpacs]{revtex4}
\usepackage{epsfig}
\usepackage{times}

\begin{document}

\title{Ne\'el order in square and triangular lattice Heisenberg
  models} 

\author{Steven~R.~White}
\author{A.~L.~Chernyshev}
\affiliation{Department of Physics and Astronomy, University of
  California, Irvine, CA 92697} 

\date{\today}

\begin{abstract}
Using examples of the square- and triangular-lattice Heisenberg models
we demonstrate that the density matrix renormalization group method (DMRG) 
can be effectively used to study magnetic ordering in two-dimensional
lattice spin models. We show that local
quantities in DMRG calculations, such as the on-site magnetization $M$, 
should be extrapolated with the truncation error, not with its square root,
as previously assumed. 
We also introduce convenient sequences of clusters,
using cylindrical boundary conditions and pinning magnetic fields,
which provide for rapidly converging finite-size scaling.
This scaling behavior on our clusters is clarified using finite-size analysis 
of the effective $\sigma$-model and finite-size
 spin-wave theory.   
The resulting greatly improved extrapolations
allow us to determine the thermodynamic
limit of $M$ for the square lattice with an error comparable to
quantum Monte Carlo. For the triangular lattice, we verify the existence
of three-sublattice magnetic order, and estimate the order parameter to 
be $M = 0.205(15)$.

\end{abstract}


\maketitle

Two-dimensional (2D) quantum lattice systems studied in condensed matter
physics can be divided into two types: those with a sign problem in
quantum Monte Carlo (QMC), and those 
without one. This is because recent developments in QMC\cite{loop,sse, troyer}
have enabled remarkably accurate large-scale
studies of the latter systems, such as the square-lattice Heisenberg
model (SLHM) \cite{Sandvik_97}.
In contrast, the former systems,
such as the triangular lattice Heisenberg model (TLHM) and other models with
geometric frustration, are often
the subject of controversy even regarding
questions of what type of order, if any, is present. 
For the TLHM, it is only recently that the rough agreement between
several theoretical\cite{swt}  
and numerical\cite{lehulier,GFMC,series} methods has made a
convincing case that the model has
three-sublattice, non-collinear 120$^\circ$ order. 
 
The density matrix renormalization group\cite{dmrg} (DMRG) is not
subject to the sign 
problem, it has an error which can be systematically decreased by
keeping more states, 
and even with modest computational effort it is extremely accurate
for one dimensional and ladder systems. For 2D systems,
the computational  
effort grows exponentially with the width.
Ameliorating this effect is the very systematic behavior of the DMRG results
versus the number of states kept, enabling the use of extrapolations 
to improve the accuracy.
The extrapolation 
of the energy versus the truncation error $\varepsilon$ 
(also known as the discarded weight) to the limit $\varepsilon \to 0$ often can
improve the accuracy of the energy by nearly an order of magnitude.
For observables other than the energy, extrapolation 
has been more problematic and is much less used.

In this Letter we show that the difficulty in extrapolating local
measurements $A$ 
is due to the incorrect assumption
that the error $\Delta A \sim \varepsilon^{1/2}$. 
In fact, the simplest way to measure local quantities within DMRG makes
$\Delta A$ {\it analytic}  in $\varepsilon$. The resulting improved
extrapolations 
greatly improve one's ability to measure order parameters in two dimensional
systems. We demonstrate this approach with a study of the SLHM and TLHM
systems. For the SLHM, the results for the on-site magnetization,
extrapolated in both  
truncation error and system size, are about as good as the best
published QMC.\cite{Sandvik_97,loopnote}  
For the TLHM, our
new results for the magnetization are comparable to
the best series expansion\cite{series} and 
GFMC\cite{GFMC} results.

Another limitation of DMRG is a large loss of accuracy if periodic boundary 
conditions (BCs) are
used lengthwise. As part of our treatment, we demonstrate an approach
using cylindrical 
BCs on $L_x\neq L_y$ clusters and pinning magnetic 
fields. We show that with an appropriate choice of the aspect ratio
$\alpha=L_x/L_y$, 
quantities such as the staggered magnetization scale much more rapidly to the 
thermodynamic limit than in widely used methods based on correlation functions 
on $L_x=L_y$ clusters with periodic BCs in both directions. 

We consider the $S=\frac{1}{2}$ Heisenberg model
\begin{equation}
H = J \sum_{\langle ij \rangle}\vec S_i \cdot \vec S_j 
\label{hamiltonian}
\end{equation}
on square and triangular lattices, where $\langle ij \rangle$ denotes
nearest neighbor 
sites, and we set $J=1$. 
We consider $L_x \times L_y$ systems with periodic BCs
in the $y$ direction, and open BCs with pinning in the $x$ direction. 
For the SLHM we consider both the standard orientation of the lattice and one 
tilted by $45^\circ$.
In all cases we apply a staggered pinning field corresponding to infinite pinning
on the edges of an  
auxiliary $(L_x+2)\times L_y$ system, e.g.
$\pm 0.5$ for the standard orientation SLHM. 
Since our DMRG program conserves total $S_z$, for the TLHM it is not possible to pin
all three sublattices simultaneously.  Instead, we only pin in the $z$ direction,  pinning one sublattice
(pointing down),   with the
other two free to rotate in a cone. Thus we expect one sublattice
in large systems 
to exhibit $\langle S_z \rangle = -M$, and the other two $+M/2$. 

We focus on the resulting onsite
magnetization $M_C = |\langle S_z \rangle|$ in the
center column of the system. 
For any fixed aspect ratio $\alpha = L_x/L_y$, $M_C$ approaches its
thermodynamic limit,  
$M_0$, as $L_x$, $L_y \to \infty$.
For $L_x \gg L_y$, the system looks more one-dimensional and we expect
$M_C$ to approach 
$M_0$ from below.
For $L_y \gg L_x$, the strong pinning dominates and we expect an
approach from above. 
We utilize intermediate values of $\alpha$
 to accelerate the convergence with system size.

\begin{figure}[t]
\includegraphics*[width=0.75\hsize,scale=1.0]{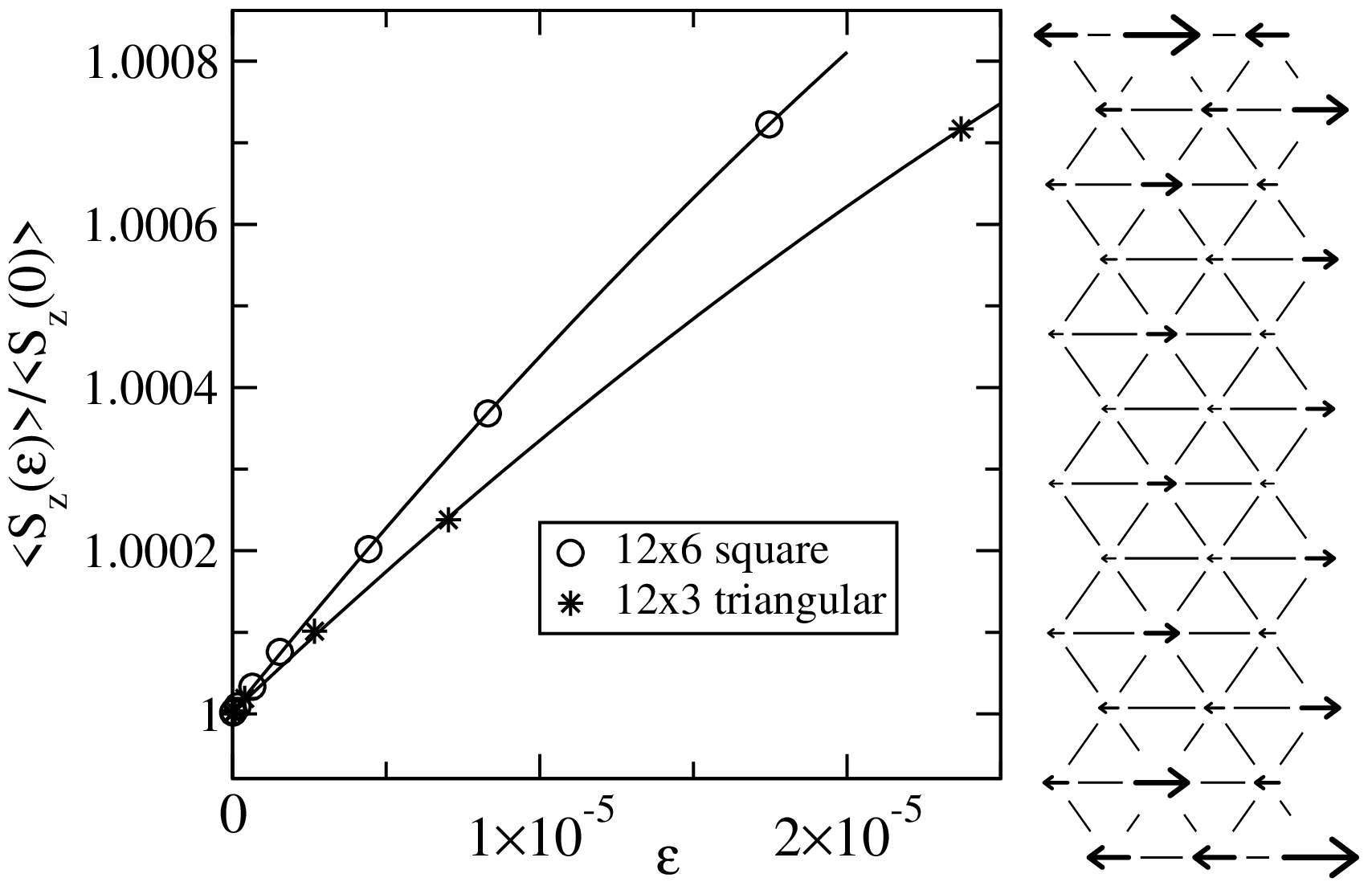}
\caption{ 
Measurements of $\langle S_z \rangle$ for a site in the middle of the
cluster with 
pinning fields applied on the ends, as a function of the truncation
error $\varepsilon$.  The 
results are normalized by the result extrapolated to $\varepsilon \to
0$. The solid lines are 
quadratic fits to the data. The
$6\sqrt{3}\times3$ triangular  
cluster, rotated $90^\circ$, is shown.
The length of the arrows is proportional to $\langle S_z \rangle$, 
and pinning fields were -0.25, -0.25, 0.5.
\label{extrap}
}
\end{figure}

First we discuss the convergence of DMRG and extrapolations in the
truncation error $\varepsilon$---the sum of the density matrix eigenvalues which are
discarded at each step.
If the truncation of density matrix states were made
starting from the exact ground state $\psi_0$, then the truncation
error and energy 
error would
vary as (to leading order) $\varepsilon  \sim \Delta E\sim |\Delta \psi|^2$, 
where $\Delta \psi = \psi - \psi_0$, and $\psi$ is the
new approximate ground state\cite{extrapnote}.
For further discussion of energy extrapolation, see
Refs. [\onlinecite{onesite,legezafath,dmrg}]. 
For measurements of an operator $\hat A$ other than the Hamiltonian, 
standard variational arguments imply an
error proportional to $\langle \Delta \psi | \hat A | \psi_0 \rangle$,
and thus $\propto \varepsilon^{1/2}$.

Consider the special situation where $\psi$ is the lowest energy state within
an incomplete basis $B$. Let $C$ be the complement of $B$.
Note that $\psi$ is an exact eigenstate in the complete basis of a
modified Hamiltonian in which the off-diagonal terms connecting $B$ and $C$ are
set to zero.  Label these coupling terms $\lambda V$, where 
$\lambda$ is an expansion parameter. Assuming $\psi$ is
close to the true ground state, $\psi_0$, 
$\lambda V\psi$ is small, and one can consider
$\lambda V$ as a small perturbation.  The leading term in $\Delta \psi$, neglecting energy denominators,
is $\propto \lambda V\psi$, which is in $C$.

Now consider a change of basis for $C$, negating each basis function. 
This sends $\lambda \to -\lambda$. Since the energy is 
independent of the change of basis, $E(\lambda)$ is even and we expect analytic
behavior for $E(\lambda^2)$.
For the exact ground state $\psi_0$, 
the change of basis switches the sign of the $C$ coefficients. 
The truncation error $\varepsilon$
is (ideally\cite{extrapnote}) 
the sum of the squares of these coefficients, and is therefore also an
even function of $\lambda$. 
Consider an operator
$\hat A$ which is block diagonal within the $B/C$ split.  Its
expectation value would also 
be independent of the change of basis, and thus an analytic function
of $\lambda^2$. 

Within DMRG, the seemingly restrictive assumption that the operator
$\hat A$ is block diagonal is  
easily satisfied for a local operator, such as $S_z$.
Consider one particular DMRG step, and consider measuring an $\hat A$  which
acts only on one or both of the central two
sites,  not part of the truncated left and right blocks. As part of
the DMRG step, 
one finds the ground state $\psi$ within the current reduced basis ($B$).
Applying $\hat A$ on $\psi$
creates a state which is exactly represented within this basis; therefore
$\hat A$ is block diagonal. At this step only a few operators can be
measured accurately, 
but as the algorithm sweeps through the lattice all local operators
can be measured.  

To utilize this analytic behavior in an extrapolation, one assumes
that successive 
sweeps, which become increasingly accurate as the number of states
kept is increased, 
corresponds to decreasing $\lambda$.
A better (but still approximate) description of the calculation is that 
the ground state is approached by taking the most significant states
out of the truncated 
basis $C$ and putting them
in $B$, not by making $\lambda$ smaller. 
We expect that in the limits of large numbers of states
kept the two types of approaches are roughly equivalent.
Then, 
both the energy and central-site operators should have polynomial (i.e.
analytic) dependence on the truncation error, and one can expect well-behaved
polynomial extrapolations.\cite{energyextrap}

In Fig. 1, we show the behavior of $\langle S_z \rangle$
as a function of $\varepsilon$ for two modest sized systems where essentially
exact results could be obtained.  The results show no signs of
nonanalytic behavior 
as $\varepsilon \to 0$, and are fit nicely 
with a quadratic form. 
We have experimented to find a reliable way to extrapolate to
$\varepsilon \to 0$, 
and have adopted
the following
simple procedure: we  utilize only the most accurate decade of data available,
and fit it with a cubic polynomial. The error bars assumed for the
purpose of the 
fit are proportional to $\varepsilon$.
The extrapolation can be checked by a fourth order
fit, or a quadratic fit over a smaller range. If these extrapolations
agree well, we take 
as a rough error estimate the empirical parameter 
0.2 times the size of the extrapolation from the last data point. If
the extrapolations 
do not agree well, we run the calculation longer if feasible, or raise
the error estimate 
substantially. 

The implications of the analytic behavior in $\varepsilon$
are significant:
local measurements for fixed $\varepsilon$ are more accurate than
previously thought, and the extrapolation $\varepsilon \to 0$ improves
results substantially and  
provides reasonable error estimates. 

We now turn to finite size effects.
Previous QMC studies of the magnetization $M$ 
have utilized correlation functions 
measured in periodic $L\times L$ systems, and extrapolation in $1/L$
for the quantity $M_0^2$.  
The leading term varies as $1/L$ with a substantial coefficient. The
expansion in $1/L$  
for the periodic $L\times L$ SLHM is
known in detail from chiral perturbation theory, allowing Sandvik to
determine $M_0\!  =\! 0.3070(3)$
using only systems up to $L\!=\!16$\cite{Sandvik_97}. For the TLHM, chiral
perturbation results are 
not available, and less robust QMC methods must be used, 
making extrapolation to $L\!\to\!\infty$ much more difficult. 
For example, Capriotti et. al. extrapolated Green's function Monte
Carlo results with 
$M^2\! \ge\! 0.13$ for $L\! \le\! 10$ down to $M_0^2\! \sim\! 0.04$ for
$L\! \to \!\infty$ to obtain $M_0\! = \!0.205(10)$. 
Other estimates\cite{swt} range as high as $M_0\!=\!0.266$.

It is known that the leading $1/L$-scaling of the order parameter 
$M$ in the 2D Heisenberg systems   
is universal and is determined by the long-wavelength spectrum of the problem, 
namely by the massless spin waves.\cite{NZ}
We have analyzed the effect of the aspect ratio $\alpha=L_x/L_y$ on
the scaling for 
pinned cylindrical and for periodic clusters using both 
finite-size scaling within an effective 
$\sigma$-model and the finite-size spin-wave theory (FSSWT). 
A key conclusion from both methods is that the coefficient in the 
$1/L$ correction to $M$ depends on $\alpha$ and, for special aspect 
ratios $\alpha_c$, vanishes, leaving corrections of order $O(1/L^2)$.
The two methods agree exactly on the values of $\alpha_c$ for nontilted and tilted square-lattice clusters: for
periodic systems, $\alpha_c = 7.0555$, 
while for cylindrical systems, for $M_C$ in the middle of the cluster,
$\alpha_c = 1.7639$, almost exactly four times smaller. 
The values of $\alpha_c$ are controlled by the cluster geometry and 
boundary conditions through the placement of the allowed 
wavevectors near the zeros of spin-wave energy: 
for periodic SLHM systems, one has
${\bf k}=(\frac{2 \pi i}{L_x},\frac{2 \pi j}{L_y})$, 
whereas for the cylindrical-pinned geometry case,
${\bf k}=(\frac{ \pi i}{L_x+1},\frac{2 \pi j}{L_y})$.  
The factor of four improvement in the aspect
ratio for the latter is due to the shift by $\frac{\pi}{L_x+1}$ away
from the ordering  vector. The effective-model  
analysis determines the $1/L$ correction term up to an unknown
factor, but the zero crossing is independent of it.

\begin{figure}[t]
\includegraphics*[width=0.9\hsize,scale=1.0]{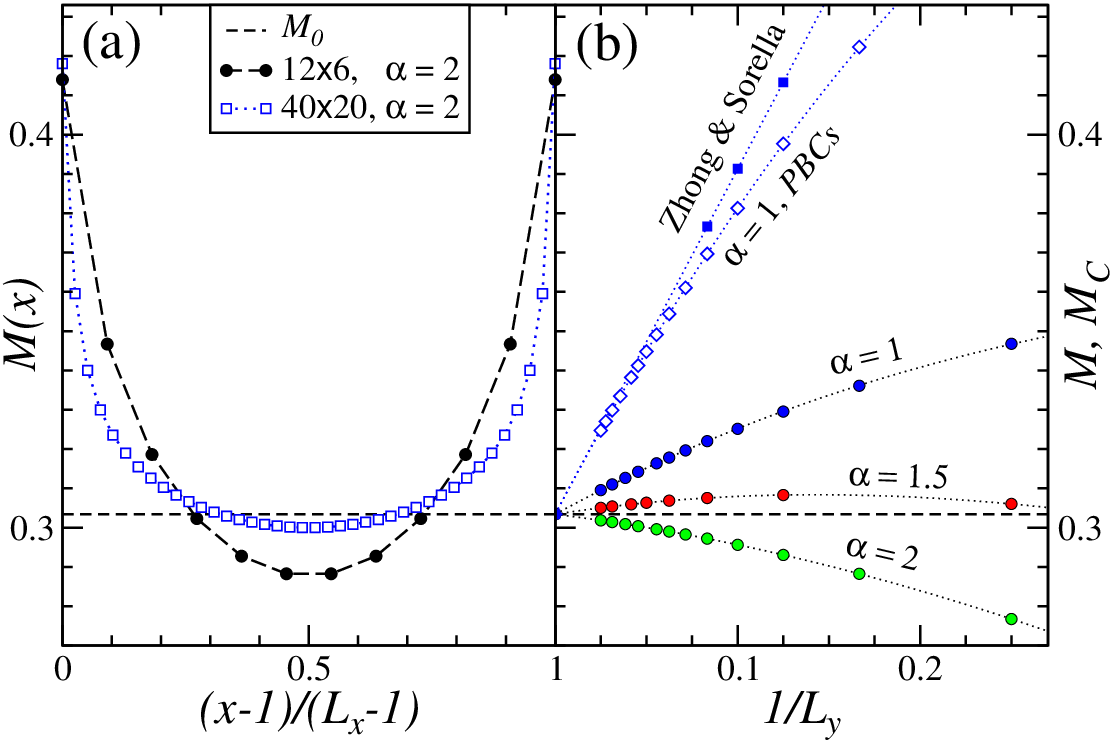}
\caption{(Color online). (a) FSSWT results for the 
SLHM showing the magnetization pattern $M(x)$ as
a function of position along the cluster $x$, for two representative clusters. 
$M_0=0.3034$ is the bulk value of the staggered magnetization within the SWT.
(b) $M_C$ ($M$) vs $1/L_y$ results for various aspect ratios $\alpha=L_x/L_y$
by FSSWT for periodic BCs (upper two sets) and 
cylindrical BCs (lower sets). In Ref. \onlinecite{Zhong_Sorella} 
$M$ was extracted from the correlation function and differs from our results 
in the higher order ($1/L^2$) terms. 
\label{fig2}
}
\end{figure}

\begin{figure}[t]
\includegraphics*[width=0.9\hsize,scale=1.0]{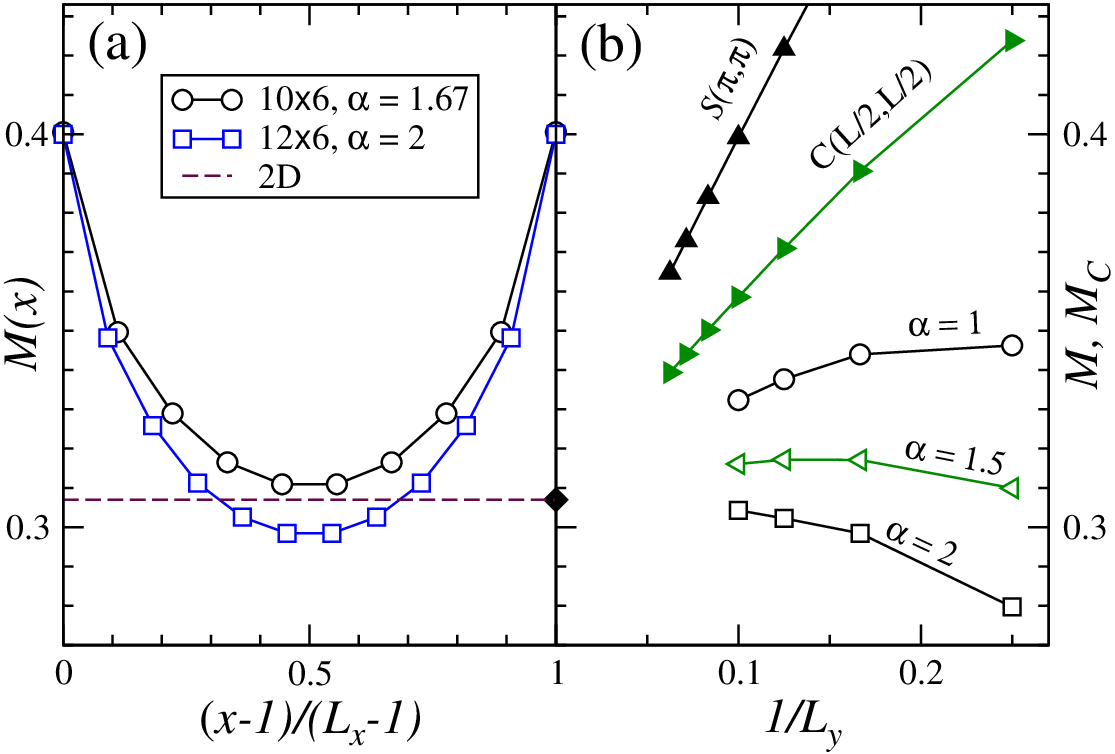}
\caption{(Color online). 
(a) $M(x)$ vs $x$ DMRG results for the SLHM, for different aspect ratios. 
The line labeled ``2D''  and the solid diamond are the QMC $L\to\infty$ extrapolated result, 
$M_0=0.3070(3)$\cite{Sandvik_97}. 
(b) $M_C$  vs $1/L_y$ results from DMRG for the SLHM. 
The upper two curves are periodic QMC $\alpha=1$ results for $M$\cite{Sandvik_97} .
\label{fig3}
}
\end{figure}

The FSSWT produces parameter-free, approximate 
results for $M=|\langle S_z \rangle|$ for all sites.  
Fig. \ref{fig2}(a) shows $M(x)$ 
vs $x$  for two representative clusters. 
Due to suppression of the long-wavelength spin fluctuations the
magnetization is enhanced  near the boundary.
The asymptotic fall-off of the magnetization away from the edge
can be shown to be $M(x)\approx M_0+a/x$,
where $a^{L=\infty}\!=\!(\pi\sqrt{8})^{-1}$. 
These FSSWT results are in a 
good agreement with the DMRG data for the SLHM in the non-tilted clusters 
shown in Fig. \ref{fig3}(a). 
One can see that already for the $L_x\times6$ clusters $M_C$
provides a good estimate of asymptotic 2D value $M_0$
when the aspect ratio is near $\alpha\!=\!2$.

Figs. \ref{fig2}(b) and \ref{fig3}(b) show $M_C$ versus $1/L_y$  for cylindrical BCs, obtained by the FSSWT and DMRG, respectively.
Also shown are
the results for the $L\times L$ systems with periodic
BCs, in Fig. \ref{fig2}(b) by FSSWT from this work and from
Ref. \onlinecite{Zhong_Sorella},  
and in Fig. \ref{fig3}(b) by QMC using standard correlation function
methods, Ref. \onlinecite{Sandvik_97}. Clearly, even for 
the same aspect ratio, the finite-size effects in the cylindrical BC
clusters are 3-4 times smaller than in the periodic systems.
The FSSWT agrees precisely with the effective theory on the
value of $\alpha_c\!=\!1.7639$ for eliminating the
leading $1/L$-term. 
This is in a good qualitative agreement with the DMRG data, but the
DMRG seem  
to indicate consistently higher values of $\alpha_c\!\approx\!1.9$. 
We have also 
performed QMC calculations\cite{alps_sse}
 for the SLHM with periodic BCs. With the largest clusters up to $20\times 160$ the
``magic'' aspect ratio is  
$\alpha_c\!\approx\!7.5$, also higher than the effective theory value
$7.0555$. While we cannot exclude a change in the behavior on larger lattice sizes, 
this seems to indicate some insufficiency of the
effective theory analysis.  

In Fig. \ref{fig3}(b) DMRG results for $M_C$ for lattices ranging up to 
$20\times10$ are shown. For the $20\times10$  system up to $m=2400$
states were kept, 
with the run taking about 40 hours single-core time on a 2.6 GHz Mac
Pro. This yielded a truncation error of order $10^{-6}$, a variational energy 
with an estimated accuracy of a part in $10^4$, 
an extrapolated energy accurate to a few parts in $10^5$,
and an uncertainty in $M_C$ of about 0.0007.

\begin{figure}[t]
\includegraphics*[width=0.9\hsize,scale=1.0]{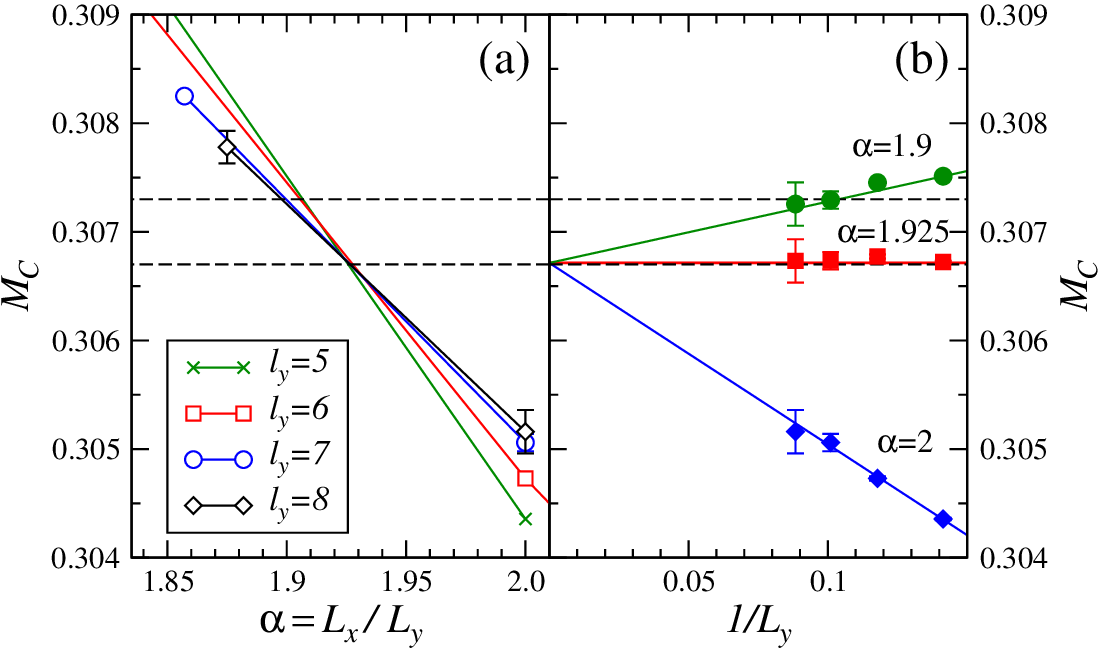}
\caption{(Color online).  DMRG results for the $45^\circ$ tilted SLHM. (a) The solid lines
  are straight segments 
connecting the discrete data points from different lattice sizes, with $L_y=l_y \sqrt{2}$. The
two dashed lines  show the bounds on the QMC result.\cite{Sandvik_97}
(b) A three parameter fit to the data from (a), as discussed in the text. 
\label{fig4}
}
\end{figure}

More accurate DMRG results can be obtained for 45$^\circ$ tilted
lattices\cite{xiang}, allowing more detailed fits. 
For example, on a  
$32/\sqrt{2}\times8\sqrt{2}$ system, the energies and
$M_C$ were roughly 2 times more accurate than for the $20\times10$
nontilted system, and  the finite size effects were smaller.
The improved behavior
comes from how DMRG sees the width of the system (the number of sites on
the boundary of the left or right block) versus the physical
dimension--the greater 
spacing by a factor of $\sqrt{2}$ in the tilted case accounts for the
improvement. 
In Fig.\ \ref{fig4}(a) we show results for $M_C$ versus $\alpha = L_x/L_y$ 
for various $L_y$ near the value $\alpha=1.925$ where the curves nearly
intersect. The intersection of such curves as $L_y \to \infty$ provides a
simple determination of both $\alpha_c$ and $M_0$. The resulting value of
$\alpha_c$, based on the available sizes, is somewhat larger than that given by
FSSWT and the continuum analysis. The values of $\alpha$ are discrete because
we have integral lattice dimensions.
Performing a least squares fit of this data to the expression
\begin{equation}
M_C(\alpha,L_y) = M_0 + a (\alpha- \alpha_c)/L_y
\label{fiteqn}
\end{equation}
we obtain $M_0 = 0.3067$, $\alpha_c =  1.9252$, and $a =  -0.1580  $. In Fig.
\ref{fig4}(b) we show a representation of this fit. The solid lines are based
on the fit; the data points for $\alpha=1.9$ and $\alpha=1.925$ are obtained
from linear extrapolation along the lines shown in (a). The result for $M_0$ is
consistent with, and of comparable
accuracy to the best QMC result.

\begin{figure}[b]
\includegraphics*[width=0.7\hsize,scale=1.0]{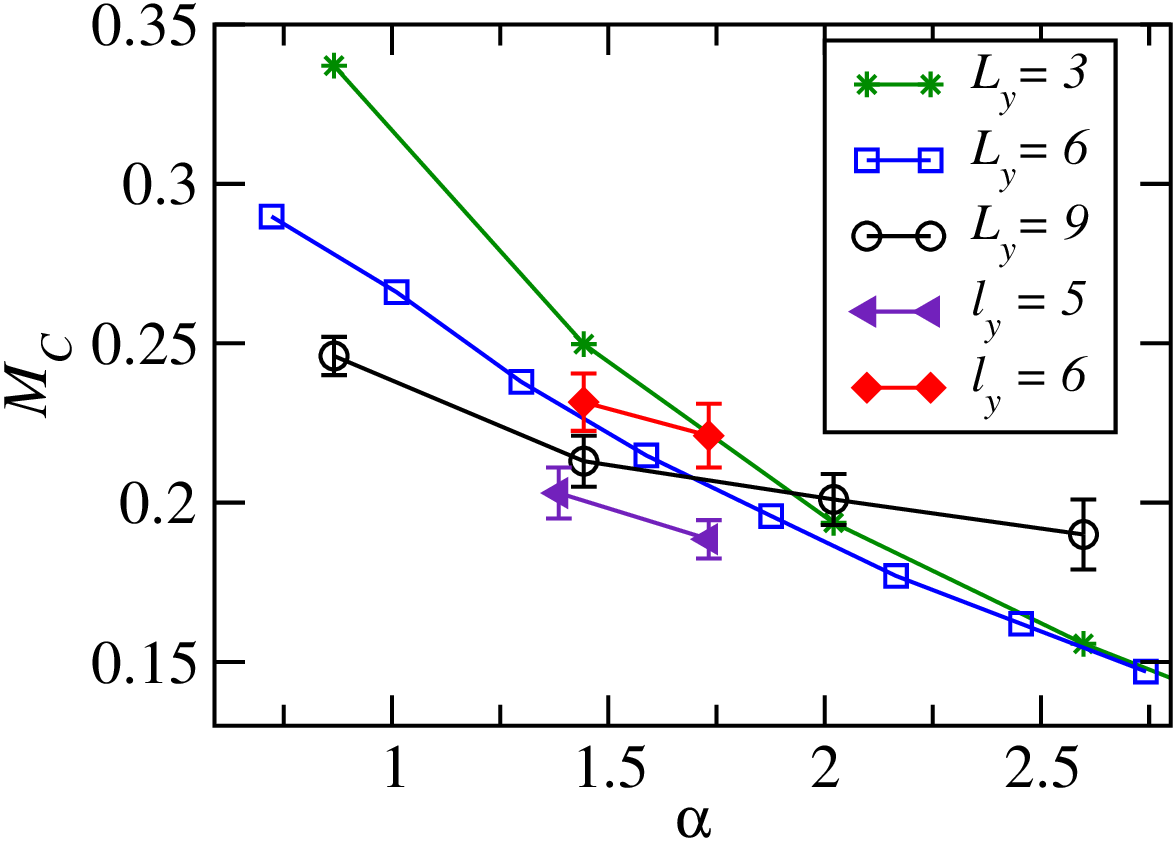}
\caption{(Color online).  $M_C$ versus aspect ratio for various widths for the TLHM, from DMRG.
The two curves labeled with $l_y$ come from clusters rotated by 90$^\circ$, with $L_y = l_y \sqrt{3}$.
\label{fig5}
}
\end{figure}

For the triangular lattice, we have studied a variety
of clusters and pinning fields; these results consistently supported that the triangular system has 
the three-sublattice 120$^\circ$ order found in other studies. The cluster
orientation
shown in Fig. 1 seems to be the most convenient and efficent for a DMRG
analysis to obtain $M_0$. 
Our DMRG results for comparable lattice sizes are only slightly less accurate
than for the  SLHM. 

Unfortunately, the finite size analysis for the TLHM is much less accurate. 
The allowed widths in the preferred geometry must
be multiples of 3, and our results for $L_y=12$ are of low
accuracy, leaving only $L_y=3, 6, 9$. Currently, we do not have comparable
analytical guidance, such as predictions for the optimal aspect ratio, for the
triangular case.  In Fig. \ref{fig5} we show results for the TLHM with this
orientation and also for lattices rotated by 90$^\circ$.
The scaling behavior appears to be quite similar to the SLHM, but with a somewhat smaller $\alpha_c \sim 1.6-1.7$. 
Assuming this behavior, we estimate $M_0 = 0.205(15)$. The results for the tilted clusters seem
to have larger finite size effects and are less useful. Our result is
consistent with recent QMC and series expansions for $M_0$ for 
the TLHM\cite{GFMC,series}.

In conclusion, we have developed improved techniques for studying ordering in 2D lattice systems
using DMRG, making DMRG competitive with QMC and series expansion methods for the 2D
Heisenberg model on square and triangular lattices.
These include proper scaling of local quantities with the discarded weight,
and the use of non-traditional cluster geometries and BCs to improve
finite-size scaling. These latter techniques can be used with other methods
besides DMRG.
We acknowledge the support of the NSF under grant DMR-0605444 (SRW), and the
DOE under grant DE-FG02-04ER46174 (ALC).


\begin{thebibliography}{99}
\bibitem{extrapnote}
In practice, one measures $\varepsilon$
relative to an approximate state, and at best $\varepsilon \propto |\Delta \psi|^2$, as  $m$ is
increased. This proportionality, which makes an 
extrapolation possible, is usually observed if the system and environment blocks
both have $m$ states, and if the energy is well-converged for this value of $m$.
\bibitem{energyextrap}  Central site operators also have a polynomial dependence 
on the error in the total energy, providing
an alternative extrapolation method, based on a separate extrapolation of the energy.
\end{thebibliography}
\end{document}